\documentclass[aps]{revtex4}
\usepackage{epsf,graphicx} 
\newcommand{\bb}{\begin{eqnarray}}
\newcommand{\ee}{\end{eqnarray}}
\begin{document}
\title{\bf Tunneling across dilaton-axion black holes}
\author{Tanwi Ghosh\footnote{E-mail: tanwi.ghosh@yahoo.co.in} and Soumitra SenGupta\footnote{E-mail: tpssg@iacs.res.in}}
\affiliation{Department of Theoretical Physics , Indian Association for the
Cultivation of Science,\\
Jadavpur, Calcutta - 700 032, India}
\vskip 5cm

\begin{abstract}
In this work we study both charged and uncharged
particles tunneling across the horizon of spherically symmetric dilaton-axion black holes using Parikh-Wilczek  tunneling formalism. 
Such black hole solutions have much significance in string theory based models. For different choices of
the dilaton and axion couplings with the electromagnetic field,
we show that the tunneling probability depends on the difference between initial and final entropies 
of the black hole. Our results, which  agrees with similar results obtained for other classes of
black holes, further confirm the usefulness of Parikh-Wilczek formalism to understand Hawking radiation.
The emission spectrum is shown to agree with a purely thermal spectrum only in the leading order. 
The modification of the proportionality factor in the area-entropy relation in the Bekenstien-Hawking formula has been determined. 
\end{abstract}
\maketitle

{{\Large {\bf Introduction}}}\\

Following Hawking's proposal \cite{hawk,hawk1,hawk2} that a black hole can radiate thermally, there has been numerous works to
investigate various aspects of this phenomenon known as 'Hawking radiation'. Such a process, which is essentially a quantum phenomenon,
predicts that a 
black hole can lose energy in the form of thermal radiation, shrink and finally evaporate.
Such thermal radiation spectrum, emitted from black holes can be  described by quantum tunneling process. According to 
this idea  once a pair of particles are created inside the horizon only the particle possessing positive energy 
comes out and the one with negative energy remains inside and lowers the mass  of the black hole. 
Recently Parikh and Wilczek \cite{pw1,pw2,pw3} proposed a formalism to describe Hawking radiation. Their approach offered 
an  explanation for the tunneling process 
of uncharged particles across the horizon of Schwarzschild and Reissner -Nordstrom black holes. 
In their explanation energy conservation along with  particle's self gravitation have been  considered.
The radiation spectrum carries the loss of mass of the black hole. By introducing  Painleve coordinate transformation, 
which is well behaved across the horizon, the metric becomes 
stationary but not static. It is shown that such a choice of coordinate is particularly useful to describe Hawking Radiation.
Recently several  works have been done to explain Hawking radiation for a large class of black holes 
using Parikh-Wilczek's tunneling 
process \cite{he,aj,aj1,ar,va1,va2,al,zh1,zh2,zh3,ji,fang,zh4,cheng,quing,arzano,wu,shu,ji1,ji2,yu,ran,gu,heng,
ya1,liu,paddy,marco,val,val1,ra1,ra2,bi1,bi2,ryan,ra3} .
In particular the method of complex paths has been developed to explain  the tunneling phenomenon in \cite{paddy}. 
Correction to the black hole entropy and it's relation to the trace anomaly has been pointed out using quantum WKB approach \cite{ra2}.
Inclusion of back reaction as well as non-commutative effects in estimating quantum tunneling has been 
developed in \cite{ra4,ra}. Moreover the connection between
the tunnelling formalism and anomaly method has been shown in \cite{ra5}. An alternative and novel formulation of tunneling formalism was
developed in \cite{ra6} where explicit complex path analysis could be avoided.\\
In some of these works particle tunneling across dilaton black hole \cite{cheng} as well as from dilaton-axion 
black hole \cite{chen,chen1,zeng1} have been investigated. The space-time metric for this class of black hole is axi-symmetric
and corresponds to
a rotating dilaton-axion black hole \cite{garcia1}. The nature of such tunneling has been explored for both scalar and 
fermionic particles. While the scalar tunneling calculations do take care of the self-gravitational interaction 
by including the back-reaction, the fermion tunneling calculations for such axion-dilaton black holes neglect the 
effects of the self-gravitation by assuming
that the energy,charge and angular momentum of the black hole remain same before and after tunneling.         

Meanwhile following string inspired black hole solutions obtained in \cite{garfinkle,gibbons,callan},
spherically symmetric Einstein-Maxwell-dilaton-axion black hole solutions were found for arbitrary dilaton and axion
couplings.\cite{sur}. For various limiting values of these couplings one arrives at 
different known black hole solutions. Several new features for these classes of black hole solutions were also obtained \cite{sur}.
Such black hole solutions have special significance in the context of
string theory and have been studied extensively.
These solutions are known to  have interesting astrophysical as well as cosmological implications \cite{say,sh,si}.
Moreover the thermodynamic properties \cite{tgssg} of such black holes are 
completely distinct from that of  rotating Einstein-Maxwell-dilaton-axion 
black hole solution considered in \cite{garcia1,chen,chen1,zeng1}.\\ 
It is therefore worthwhile to explore the character of Hawking radiation for  both massless and 
massive as well as  uncharged and charged  particles tunneling across such
spherically symmetric dilaton-axion black holes. In this work we consider scalar particle tunneling to incorporate the back-reaction
and re-examine this in the light of Parikh-Wilczek tunneling formalism for these entirely new classes of black holes.
We show that the tunneling rate for these class of generalised spherically symmetric black holes once again is given by
the generic result found by Parikh and Wilczek and is independent of the mass of the tunneling particle. It was further shown that 
in the leading order
the emission spectrum exhibits a thermal character though a deviation is found because of  higher order effects. 
Finally 
we have determined the semiclassical corrections \cite{vagenas} to the proportionality factor in the 
area-entropy relation over it's usual value of $1/4$.    

We begin with the action describing dilaton-axion black holes as follows \cite{sur}:\\

\begin{eqnarray}
S=\int d^4x\sqrt{-g}[\frac{1}{2\kappa}(R-\frac{1}{2}\partial_{\mu}\varphi\partial^{\mu}\varphi
-\frac{1}{2}e^{2a\varphi}\partial_{\mu}\zeta\partial^{\mu}\zeta)-e^{-a\varphi}F_{\mu\nu}F^{\mu\nu}
-b\zeta F_{\mu\nu}*F^{\mu\nu}]
\end{eqnarray} 
where '$a$' and '$b$' are coupling parameters of dilaton field $\varphi$ and axion 
field $\zeta$ respectively with the Maxwell field tensor $F_{\mu\nu}$.The parameter '$a$' which is real and 
non-negative originates from Gibbons-Maeda's dilaton black holes \cite{gibbons}. The axion-electromagnetic coupling 
parameter 'b' appears in the four dimensional effective theory of string theory where the axion in four dimension
is connected to the third rank antisymmetric field strength tensor of the second rank anti-symmetric closed
string mode $B_{\mu\nu}$ ( called Kalb-Ramond field) via a duality relation.
The Kalb-Ramond three tensor field $H_{\mu\nu\lambda}$ has the usual interpretation of torsion in spacetime. 
The coupling parameter '$b$' whose value in general depends on the nature of compactification and on the vacuum expectation values
of the moduli fields , is taken here as a free parameter. This enables us to keep our study quite general.\\

{{\Large {\bf Chargeless Particle Tunneling From Dilaton-Axion Black hole }}}.\\

Considering above generic Einstein-Maxwell-Kalb-Ramond action, a static spherically symmetric black 
hole solution for arbitrary coupling parameters is given by the metric\cite{sur},
\begin{eqnarray}
ds^2= -\frac{(r-r_+)(r-r_-)}{(r-r_0)^{2-2n}(r+r_0)^{2n}}dt_s^2
+\frac{(r-r_0)^{2-2n}(r+r_0)^{2n}}{(r-r_+)(r-r_-)}dr^2+\frac{(r+r_0)^{2n}}{(r-r_0)^{(2n-2)}}
[d\theta^2+sin^2\theta d\phi^2]    
\end{eqnarray}\\
Arbitrary coupling parameters a and b appear through the following equations:
$\gamma=4q_m^2e^{-a\varphi}+4(q_e-q_mb\zeta)^2e^{a\varphi}$; $\psi'^2=\varphi'^2+e^{2a\varphi}\zeta'^2$.  
$\psi$ and $\gamma$ respectively are the effective scalar and effective 
coupling in the theory while the scalar-shielded electric and magnetic charges are $q_e$ and $q_m$.

The non-vanishing electromagnetic field components are given by, 
\begin{eqnarray} 
F_{tr}=\frac{(q_e-q_mb\zeta)e^{a\varphi}}{(r-r_0)^{2(1-n)}(r+r_0)^{2n}}dt dr
\end{eqnarray}
and,
\begin{eqnarray}
F_{\theta\phi}=q_m sin{\theta}d\theta d\phi
\end{eqnarray}
The black hole horizons are located at,
\begin{eqnarray}
r_{\pm} = m_0 \pm\sqrt{m_0^2+r_0^2-\frac{1}{8}(\frac{K_1}{n}+\frac{K_2}{1-n})} 
\end{eqnarray}
where,
$r_0=\frac{1}{16m_o}(\frac{K_1}{n}-\frac{K_2}{1-n})$ ; $m_0= m-(2n-1)r_0$ ; $ m = \frac{1}{16r_0}(\frac{K_1}{n}-\frac{K_2}{1-n})+(2n-1)r_0$\\
with $K_1 = 4n[4r_0^2 +2kr_0(r_++r_-)+k^2r_+r_-]$ ; $K_2=4(1-n)r_+r_-$;  and $0<n<1$.\\ 
Here $m$ is the mass of the black hole. The value of $k$ is one for 
asymptotically flat case.\\ 
Using the usual expressions for area and Hawking temperature for spherically symmetric black holes, we determine these quantities for
the above dilaton-axion black hole as, 
\begin{eqnarray}
T=\frac{(r_+-r_-)}{[4\pi(r_0+r_+)^{2n}(r_+-r_0)^{2-2n}]}
\end{eqnarray}
\begin{eqnarray}
A=4\pi\frac{(r_++r_0)^{2n}}{(r_+-r_0)^{(2n-2)}}
\end{eqnarray}

To explore the tunneling phenomenon from such black holes we follow the  method proposed by Parikh and Wilczek \cite{pw1,pw2,pw3}
as well as the general analysis shown in \cite{ra} in the context of a 
static spherically symmetric metric.
We use Painleve-Gullstrand coordinate in which the  black hole spacetime can be written as \cite{he}.
\begin{eqnarray}
dt_s=dt-G(r)dr
\end{eqnarray}
$G(r)$ satisfies the condition $f^{-1}(m,Q,r)-f(m,Q,r)G^2(r)=1$,
where $f(m,Q,r)=\frac{(r-r_+)(r-r_-)}{(r-r_0)^{2-2n}(r+r_0)^{2n}}$.\\
The corresponding black hole metric is given as,
\begin{eqnarray}
ds^2=-f(m,Q,r)dt^2+2\sqrt{1-f(m,Q,r)}dtdr+dr^2+h(m,Q,r)(d\theta^2+sin^2{\theta}d\phi^2)
\end{eqnarray}
where $h=\frac{(r+r_0)^{2n}}{(r-r_0)^{(2n-2)}}$

One of the the advantage of Painleve coordinate transformations is 
that the solution becomes well-behaved without any singularity at the horizon.
Coordinate time $t$ coincides with the local proper time of an observer falling freely across a radial trajectory from infinity. 
Measurements performed on constant-time slices are therefore same as that on flat Euclidean spacetime.
An observer at infinity can not distinguish between Painleve coordinates and static black hole coordinates.
We further mention that the Landau's
condition \cite{lan} of coordinate clock synchronization holds during particle tunneling
across the horizon of the black hole.\\
In this case  we consider massive uncharged particles tunneling
across the horizon. The radial motion of the particle is given by, 
\begin{eqnarray}
\dot{r}=-\frac{g_{00}}{2g_{01}}=\frac{f(m,Q,r)}{2\sqrt{1-f(m,Q,r)}}
 \end{eqnarray}
Substituting for $f(m,Q,r)$,
\begin{eqnarray}
\dot{r} =\frac{(r-r_+)(r-r_-)}{2(r-r_0)^{2-2n}(r+r_0)^{2n}}\frac{1}{\sqrt{1-\frac{(r-r_+)(r-r_-)}{(r-r_0)^{2-2n}(r+r_0)^{2n}}}}
\end{eqnarray}
Along with the above mentioned property of Painleve coordinate system,the 
event horizon coincides with the infinite redshift surface in the dragged Painleve coordinate system.
This enables us to use WKB approximation to find out the expression for tunneling probability $\Gamma$ for an 
outgoing positive energy particle as,\\

\begin{eqnarray}
\Gamma\approx e^{-2Im I}
\end{eqnarray}
The imaginary part of the action $I$ can be written as
\begin{eqnarray}
Im I = Im \int_{r_i}^{r_f}(P_r \dot{r})dt\\
     =Im\int_{r_i}^{r_f}\int_0^{P_r}(\dot{r}dP_r')\frac{dr}{\dot{r}}
\end{eqnarray}
where $P_r$ is the conjugate momenta corresponding to r. The outer event horizons before and after the emission of the uncharged 
particle are located respectively at,
$r_i = m_o +\sqrt{m_o^2+r_o^2-\frac{1}{8}(\frac{K_1}{n}+\frac{K_2}{1-n})}$  
and
$r_f=m_o -\omega +\sqrt{(m_o - \omega)^2 + r_{0f}^2-\frac{1}{8}(\frac{K_1}{n}+\frac{K_2}{1-n})}$\\
$r_{0f}$ is the modified value of $r_o$ after the emission and can be obtained by replacing $m_0$ by $m_0 - \omega$.
The Hamilton's equation of motion is given as,
\begin{eqnarray}
\dot{r}=\frac{dH}{dP_r}=\frac{d(m-\omega)}{dP_r}
\end{eqnarray}
Here we take into account particle's self gravity effect\cite{kraus1,kraus2}.
It may be observed that while hole's mass can fluctuate but to conserve energy, the  
energy of the spacetime remains fixed.
Thus when a particle of energy $\omega$ is radiated, the black hole mass reduces to $m-\omega$.
The corresponding tunneling probability can be derived as:
\begin{eqnarray}
\nonumber Im I= Im\int_{r_i}^{r_f}\int_m^{m-\omega}d(m-\omega)\frac{dr}{\dot{r}}\\
\end{eqnarray}    
Substituting for $\dot{r}$ from equation(11),
\begin{eqnarray}
\nonumber Im I =Im\int_{r_i}^{r_f}\int_m^{m-\omega}dm'dr\frac{2(r-r_0')^{2-2n}(r+r_0')^{2n}\sqrt{1-\frac{(r-r_+')(r-r_-')}{(r-r_0')^{2-2n}(r+r_0')^{2n}}}}{(r-r_+')(r-r_-')}\\
\end{eqnarray}
Evaluating the integral around the pole we obtain,
\begin{eqnarray}
Im I =-2\pi\int_m^{m-\omega}dm'\frac{(r_0'+r_+')^{2n}(r_+'-r_0')^{2-2n}}{(r_+'-r_-')}
\end{eqnarray}
It may be noted that each of the parameters $m$, $r_{+}$, $r_{-}$ etc changes from their initial value before tunneling to the 
corresponding final value after tunneling. For example the mass changes from $ m = m $ to the final value for $ m = m - \omega$ 
due to tunneling. Therefore during tunneling they have been replaced by variable parameters with a prime on them inside the integrals. 
Also we  have defined $m'=m-\omega'$ and $m_o'= (m-\omega)'-(2n-1)r_0'$.\\
Using the expression of Hawking temperature from equation (6), the first law of black hole thermodynamics can be expressed as,
\begin{eqnarray}
\frac{dm'}{\frac{(r_+'-r_-')}{[4\pi(r_0'+r_+')^{2n}(r_+'-r_0')^{2-2n}]}}= dS'
\end{eqnarray}
Equation (18) therefore becomes,
\begin{eqnarray}
Im I=-\frac{1}{2}\Delta S'
\end{eqnarray}
In terms of the entropy S, the tunneling probability therefore becomes,
\begin{eqnarray}
\Gamma=e^{\Delta S_{bh}}
\end{eqnarray}
This exhibits the relation between the tunneling probability and the entropy difference of black hole before and after particle emission.\\
Let us now focus into the tunneling of a massless particle across the black hole.
In this case one needs to consider the radial null geodesic equation which is given as,
\begin{eqnarray}
-f+2\sqrt{1-f}\dot{r}+\dot{r}^2=0
\end{eqnarray}
The expression for the radial velocity therefore becomes,
\begin{eqnarray}
\dot{r}= \pm 1-\sqrt{1-\frac{(r-r_+)(r-r_-)}{(r-r_0)^{2-2n}(r+r_0)^{2n}}}
\end{eqnarray}
where $\pm$ sign are due to the outgoing and the incoming particles respectively.
Substituting $\dot{r}$ in equation (16),deforming the integral around the pole at $r=r_+$,we get
\begin{eqnarray}
Im I = -2\pi\int_m^{m-\omega}dm'\frac {(r_0'+r_+')^{2n}(r_+'-r_0')^{2-2n}}{(r_+'-r_-')}
     =-\frac{1}{2}\int dS' =-\frac{1}{2}\Delta S'
\end{eqnarray}
Thus the tunneling probability $\Gamma$ once again becomes,
\begin{eqnarray}
\Gamma=e^{\Delta S_{bh}}
\end{eqnarray}\\
We would now obtain the corrected coefficient of Bekenstein-Hawking area-entropy law over the usual factor of $1/4$.
Here ,
\begin{eqnarray}
\Delta S_{bh}=S_{bh}(\omega,m)-S_{bh}(m)
\end{eqnarray}
Due to self gravity effect,
\begin{eqnarray}
S_{bh}(\omega,m)=S_0+\frac{\pi (r_{+out}+r_{0 out})^{2n}}{(r_{+out}-r_{0out})^{2n-2}}
\end{eqnarray}
where $S_0$ can be found in terms of Bekenstein-Hawking entropy $S_{BH}=\pi r_{+}^2$ as,
\begin{eqnarray}
S_0=S_{BH}-\pi r_{+in}^2\frac{(1+\frac{r_{0in}}{r_{+in}})^{2n}}{(1-\frac{r_{0in}}{r_{+in}})^{2n-2}}
\end{eqnarray}
Expanding $r_{+out}$ in order of $\omega$  we get,
\begin{eqnarray}
r_{+ out}=r_{+ in}(1-\epsilon\omega)
\end{eqnarray}
where parameter $\epsilon=\frac{1}{m}$\\
Substituting $S_0$, $r_{+out}$, $r_{+in}$ and  $r_{0out}$,we get
\begin{eqnarray}
S_{bh}(\omega,m)=S_{BH}[1-\epsilon\omega C]
\end{eqnarray}
where 
$C=(2-(2-4n)\frac{(\frac{K_1}{n}-\frac{K_2}{1-n})}{4m^2(1+\sqrt{1-\frac{(2n-1)(\frac{K_1}{n}-\frac{K_2}{1-n})}{4m^2}})^2}
\frac{1}{[1+(1+\frac{(\frac{K_1}{n}-\frac{K_2}{1-n})^2}{16(m+\sqrt{m^2-\frac{(2n-1)(\frac{K_1}{n}-\frac{K_2}{1-n})}{4}})^4}-\frac{(\frac{K_1}{n}+\frac{K_2}{1-n})}{2(m+\sqrt{m^2-\frac{(2n-1)(\frac{K_1}{n}-\frac{K_2}{1-n})}{4}})^2})^{\frac{1}{2}}]})$
and the constants $K_1$ and $K_2$ are given in equation (5) in terms of t the black hole horizons $r_{\pm}$.

As  $\epsilon \sim 1/m$, $\omega < m $ and $C \sim 1/m^2$ therefore $S_{bh} < S_{BH}$. This result is in agreement with the 
semi-classical correction to the area-entropy preportionality factor found in \cite{vagenas}.\\

{{\Large {\bf Charged Particle Tunneling From Dilaton-Axion Black hole }}}.\\

{{\Large {\bf Case-I}}\\

We now focus into the tunneling of massive charged particles across dilaton-axion black hole when
dilaton coupling $a$ and axion coupling $b$ are same i.e  $|a|=|b|$. In this scenario, 
we have to consider the conservation of electric charge in addition to the conservation of energy.
To preserve energy as well as electric charge,we will fix the total mass and  charge of the spacetime and only vary those of 
the black hole. Consider a particle having energy $\omega$ and electric charge q  
tunneling across an electrically charged  dilaton-axion black hole. The space-time metric is described by \cite{sur},\\

\begin{eqnarray}
ds^2=-(1-\frac{2m_0}{r})(1-\frac{2r_0}{r})^{\frac{(1-a^2)}{(1+a^2)}}dt_s^2+(1-\frac{2m_0}{r})^{-1}(1-\frac{2r_0}{r})^{\frac{(a^2-1)}{(1+a^2)}}dr^2+r^2(1-\frac{2r_0}{r})^{\frac{2a^2}{(a^2+1)}}d\Omega^2
\end{eqnarray}
Where parameters $m_0$,$r_0$ ,the black hole mass $m$  total charge $Q$,electric charge $Q_e$ and
magnetic charge $Q_m$ are related as:\\
\begin{eqnarray}
r_0=\frac{(1+a^2)Q^2e^{-a\varphi_0}}{4m_0}\\
m_0=m-\frac{(1-a^2)}{(1+a^2)}r_0\\
Q^2=Q_e^2+Q_m^2
\end{eqnarray}
$\varphi_0$ is the asymptotic value of the scalar field. These solutions represent black holes with it's 
horizon  located at $r=2m_0$. The solution however has a curvature singularity at $r=2r_0$.\\
The area A and temperature T of this black hole respectively are given as,
\begin{eqnarray}
A=4\pi r_+^2(1-\frac{2r_0}{r_+})^{\frac{2a^2}{(a^2+1)}} 
\end{eqnarray}
and
\begin{eqnarray}
T=\frac{1}{4\pi r_+}(1-\frac{2r_0}{r_+})^{\frac{(1-a^2)}{(1+a^2)}}
\end{eqnarray}
Once again we consider the following Painleve-type coordinate transformation:
\begin{eqnarray}
dt_s=dt-G(r)dr
\end{eqnarray}
such that $f(m,Q,r)^{-1}-f(m,Q,r)G^2=1$ with $f(m,Q,r)=(1-\frac{2m_0}{r})(1-\frac{2r_0}{r})^{\frac{(1-a^2)}{(1+a^2)}}$

After this transformation the line element ( see equ. 26 ) can be written as
\begin{eqnarray}
ds^2=-f(m,Q,r)dt^2+2\sqrt{1-f(m,Q,r)}dtdr+dr^2+h(m,Q,r)(d\theta^2+sin^2{\theta}d\phi^2)
\end{eqnarray}
where $h(m,Q,r)=r^2(1-\frac{2r_0}{r})^{\frac{2a^2}{(a^2+1)}}$.
The non-zero component of the vector potential is given as,
\begin{eqnarray}
A_t=\frac{Q_e e^{-a\varphi_0}}{r}
\end{eqnarray}
After the Painleve coordinate transformation none of the components of the metric
or the inverse metric diverge. Following de-Broglie's hypothesis phase velocity $v_p$ and group 
velocity $v_g$ will be related through $v_p=\frac{1}{2}v_g$. As before Landau's coordinate clock synchronization now implies,
\begin{eqnarray}
dt=-\frac{g_{tr}}{g_{tt}}dr_c (d\theta=0,d\phi=0)
\end{eqnarray}
The radial velocity of the particle is obtained as,
\begin{eqnarray}
\dot{r}=-\frac{g_{tt}}{2g_{01}}=\frac{f(m,Q,r)}{2\sqrt{1-f(m,Q,r)}}
\end{eqnarray}
It should be remembered that during charged particle tunneling mass m and 
charge Q should be replaced by $m-\omega$ and $Q-q$. Here in tunneling process,
we need to consider matter-gravity system consisting of black hole and the electromagnetic field outside the black hole.
The Lagrangian of electromagnetic field is given as  $L_{elec}=-\frac{1}{4\pi}F_{\mu\nu}F^{\mu\nu}$ .
Total action of the matter-gravity system can be written as,
\begin{eqnarray}
I=\int_{t_i}^{t_f}(P_t+P_r \dot{r}-P_{A_t}\dot{A_t})dt
\end{eqnarray}
where $P_t$,$P_r$ and $P_{A_t}$ are the conjugate momenta corresponding to $(t,r,A_t)$. $t_i$ and $t_f$ are the 
Painleve coordinate times before and after the particle tunneling.
Eliminating $A_t$ from the Lagrangian, action can be written as,
\begin{eqnarray}
I=\int_{r_i}^{r_f}\int_{0,0}^{P_r,P_{A_t}}(dP'_r-\frac{\dot{A'_t}}{\dot{r}}dP'_{A_{t}})dr
\end{eqnarray}
Considering particle's self gravitational effect and using Hamilton's equations,
\begin{eqnarray}
\dot{r}=\frac{dm'}{dP'_r}\\
\dot{A_t'}=\frac{dE'_{Q}}{dP'_{A_t}}
\end{eqnarray}
the action can be written as
\begin{eqnarray}
I=\int_{r_i}^{r_f}\int_{m,E_{Q}}^{m-\omega,E_{Q-q}}\frac{dm'-dE'_{Q}}{\dot{r}}dr
=\int_{r_i}^{r_f}\int_{m,Q}^{m-\omega,Q-q}(dm'-\frac{Q'dQ'e^{-a\varphi_0}}{r})\frac{dr}{\dot{r}}
\end{eqnarray}
where $E'_Q$ represents the energy of electro-magnetic field,$r_i=2m_0$ and $r_f=2(m_0 - \omega)$ are 
the horizon locations before and after the 
emission of charged particle and in the dragged 
frame the expressions of $r_0'$ and $m_0'$ can be written 
respectively as 
$m_0'=m'-\frac{(1-a^2)}{(1+a^2)}r_0'=(m-\omega')-\frac{(1-a^2)}{(1+a^2)}r_0'$ with $r_0'=\frac{(1+a^2)(Q-q)^2e^{-a\varphi_0}}{4m_0'}$.
We will now substitute $\dot{r}$ in the action to 
find out the emission rate. Using WKB approximation the tunneling probability of the particle can be expressed as
\begin{eqnarray}
\Gamma\approx e^{-2Im{I}}
\end{eqnarray}
where 
\begin{eqnarray}
\nonumber Im{I}=Im\int_{m,Q}^{m-\omega,Q-q}\int_{r_i}^{r_f}(dm'-\frac{Q'dQ'e^{-a\varphi_0}}{r})dr
\frac{2\sqrt{1-(1-\frac{2m_0}{r})(1-\frac{2r_0}{r})^{\frac{(1-a^2)}{(1+a^2)}}}}
{(1-\frac{2m_0}{r})(1-\frac{2r_0}{r})^{\frac{(1-a^2)}{(1+a^2)}}}\\
\end{eqnarray}
Calculating residue at $r=r_+=2m_0$,
\begin{eqnarray}
Im{I}=-\int 2\pi r_+'(1-\frac{2r_0'}{r_+'})^{\frac{(a^2-1)}{(a^2+1)}}(dm'-\frac{Q'dQ'e^{-a\varphi_0}}{r_+'})
\end{eqnarray}
Using the expression for Hawking temperature from equations (31) the first law of black hole mechanics can be written as,
\begin{eqnarray}
\frac{(dm'-\frac{Q'dQ'e^{-a\varphi_0}}{r_+'})}{ \frac{1}{4\pi r_+'}(1-\frac{2r_0'}{r_+'})^{\frac{(1-a^2)}{(1+a^2)}}} =dS'
\end{eqnarray}
The equation (44) therefore becomes,
\begin{eqnarray}
Im{I}=-\int\frac{1}{2}dS'=-\frac{1}{2}\Delta S
\end{eqnarray}
In an alternative approach using the expression for the horizon area from equation(30), the expression of entropy is obtained as 
$S=\pi r_+^2(1-\frac{2r_0}{r_+})^{\frac{2a^2}{(a^2+1)}}$. 
The change in entropy in dragged frame therefore can be expressed as, 
\begin{eqnarray}
dS'=4\pi r_+'(1-\frac{2r_0'}{r_+'})^{\frac{(a^2-1)}{(a^2+1)}}[dm'-\frac{Q'dQ'e^{-a\varphi_0}}{r_+'}]
\end{eqnarray}
Thus tunneling rate becomes,
\begin{eqnarray}
\Gamma\approx e^{\Delta S_{bh}}
\end{eqnarray}
This confirms the validity of Parikh-Wilczek tunneling formalism for this class of black holes also.\\

To explore the tunneling characteristics of massless uncharged particles for this class of black holes 
we use the appropriate expression of $\dot{r}$ for outgoing massless and uncharged particle:
\begin{eqnarray}
\dot{r}=1-\sqrt{1-(1-\frac{2m_0}{r})(1-\frac{2r_0}{r})^{\frac{(1-a^2)}{(1+a^2)}}}
\end{eqnarray}
It is now easy to show that,
\begin{eqnarray}
Im{I}=-\int_{m}^{m-\omega}\int_{r_i}^{r_f}2\pi r_+'(1-\frac{2r_0'}{r_+'})^{\frac{(a^2-1)}{(a^2+1)}}dm'
\end{eqnarray}
Using first law of black hole thermodynamics,
\begin{eqnarray}
dS'=4\pi r_+'(1-\frac{2r_0'}{r_+'})^{\frac{(a^2-1)}{(a^2+1)}}[dm']
\end{eqnarray}
the tunneling rate becomes
\begin{eqnarray}
\Gamma\approx e^{\Delta S_{bh}}
\end{eqnarray}\\

We now obtain the correction to the coeficient in the Bekenstein-Hawking area-entropy law over the usual factor of $1/4$.
We have already shown that for tunneling of both massless and massive charged particles the emission rate is proportional
to the exponent of $\Delta S_{bh}$,where 
\begin{eqnarray}
\Delta S_{bh} = S_{bh}(\omega,m)-S_{bh}(m)
\end{eqnarray}
Due to self gravitational effect the modified entropy can be expressed as,
\begin{eqnarray}
S_{bh}(\omega,m)=S_0 +\pi r_{+ out}^2(1-\frac{2r_{0 out}}{r_{+ out}})^{\frac{2a^2}{(a^2+1)}}
\end{eqnarray}
To determine the  arbitrary constant $S_0$ we use Bekenstein-Hawking area law for  black hole entropy $S_{BH} = \pi r_+^2$. 
Using this $S_{BH}$, $S_0$ can be expressed as,
\begin{eqnarray}
S_0 = S_{BH}-S_{BH}(1-\frac{2r_{0 in}}{r_{+ in}})^{\frac{2a^2}{(a^2+1)}}
\end{eqnarray}
Now substituting $S_{BH}$ and $S_0$, $S_{bh}(\omega,m)$ will be
\begin{eqnarray}
S_{bh}(\omega,m)= S_{BH}+ [\pi r_{+ out}^2(1-\frac{2r_{0 out}}{r_{+ out}})^{\frac{2a^2}{(a^2+1)}}-
\pi r_{+ in}^2(1-\frac{2r_{0 in}}{r_{+ in}})^{\frac{2a^2}{(a^2+1)}}]
\end{eqnarray}
To examine the modified entropy i.e whether it is larger or smaller than $S_{BH}$, we now expand $r_{+ out}$ w.r.t first order
in $\omega$ and get,
\begin{eqnarray}
r_{+ out}=r_{+ in}(1-\epsilon\omega)
\end{eqnarray}
where parameter $\epsilon=\frac{1}{m(1+\sqrt{1-
\frac{(1-a^2)Q^2e^{-a\varphi_0}}{m^2}})}$.\\
Substituting $r_{+ out}$ in the 
expression of $S_{bh}(\omega,m)$, we get keeping terms upto first order in $\omega$ as,
\begin{eqnarray}
S_{bh}(\omega,m)= S_{BH}[1-\epsilon\omega(2-\frac{2a^2(a^2+1)(Q-q)^2e^{-a\varphi_0}}{m^2(1+\sqrt{1-
\frac{(1-a^2)Q^2e^{-a\varphi_0}}{m^2}})^2})]
\end{eqnarray}
For zeroth order in $\omega$, equality between $S_{bh}(\omega,m)$ and $S_{BH}$ holds.\\ 
However due to tunneling the modified entropy takes the
form,
\begin{eqnarray}
S_{bh}(\omega,m)= \alpha_1 S_{BH}=\alpha_1\frac{A_{{BH}}}{4}
\end{eqnarray}
where $\alpha_1=[1-\epsilon\omega(2-\frac{2a^2(a^2+1)(Q-q)^2e^{-a\varphi_0}}{m^2(1+\sqrt{1-
\frac{(1-a^2)Q^2e^{-a\varphi_0}}{m^2}})^2})]$.\\ 
Now since $(2-\frac{2a^2(a^2+1)(Q-q)^2e^{-a\varphi_0}}{m^2(1+\sqrt{1-
\frac{(1-a^2)Q^2e^{-a\varphi_0}}{m^2}})^2})<2$ for non-zero value of $a$, therefore 
$\alpha_1< 1$. 
This implies,
\begin{eqnarray}
S_{bh}(\omega,m)\leq S_{BH}
\end{eqnarray}
This in turn indicates that the entropy of this kind of black hole is less than the usual Bekenstein-Hawking entropy
$S_{BH} = \frac{A_{{BH}}}{4}$. \\

{{\Large {\bf Case-II}}\\

We finally consider dilaton-axion black hole where dilaton coupling $a$ and axion 
coupling $b$ are related through $|a|\neq|b|$ and $b<<1$. This kind of black hole can be 
expressed through the following spacetime metric \cite{sur},
\begin{eqnarray}
ds^2=\frac{-(r-r_+)(r-r_-)}{(r^2-r_0^2)}dt^2+\frac{(r^2-r_0^2)}{(r-r_+)(r-r_-)}dr^2
+(r^2-r_0^2)(d\theta^2+sin^2{\theta}d\phi^2)
\end{eqnarray}
where 
\begin{eqnarray}
r\pm=m\pm\sqrt{m^2+r_0^2-(Q_e^2+Q_m^2)e^{-\varphi_0}}
\end{eqnarray}
are the event horizons of this kind of black hole 
and $Q^2=(Q_e^2+Q_m^2)$.$\varphi_0$ is the asymptotic value of the 
scalar field.Maxwell field $F_{01}$ is expressed by $F_{01}=\frac{Q_e e^{-\varphi_0}}{(r+r_0)}$ where 
constant parameter $r_0=\frac{(Q_e^2-Q_m^2)e^{-\varphi_0}}{2m}$.m,$Q_e$ and $Q_m$ are the mass,
electric charge and magnetic charge of the black hole.Area A and temperature T can be expressed as,
\begin{eqnarray}
A=4\pi(r_+^2-r_0^2)
\end{eqnarray}
and
\begin{eqnarray}
T=\frac{1}{\beta}=\frac{(r_+-r_-)}{4\pi(r_+^2-r_0^2)}
\end{eqnarray}
Non vanishing component of vector potential is given by,
\begin{eqnarray}
A_t=\frac{Q_e e^{-\varphi_0}}{(r+r_0)}
\end{eqnarray}
Using the similar argument and Painleve coordinate transformation as in Case-I, equation (66) can be written as 
\begin{eqnarray}
ds^2=-f(m,Q,r)dt^2+2\sqrt{1-f(m,Q,r)}dtdr+dr^2+h(m,Q,r)(d\theta^2+sin^2{\theta}d\phi^2)
\end{eqnarray}
where $f=\frac{(r-r_+)(r-r_-)}{(r^2-r_0^2)}$ and $h=(r^2-r_0^2)$.
We can find the radial velocity of massive charged particle tunneling through this black hole horizon as
\begin{eqnarray}
\dot{r}=-\frac{g_{tt}}{2 g_{tr}}=\frac{f}{2\sqrt{1-f}}
\end{eqnarray}
Substituting $\dot{r}$ in 
equation (41), ImI becomes,
\begin{eqnarray}
\nonumber Im{I}=Im\int_{m,Q}^{m-\omega,Q-q}\int_{r_i}^{r_f}(dm'-\frac{Q'dQ'e^{-\varphi_0}}{r})dr
\frac{2(r^2-r_0^2)\sqrt{1-\frac{(r-r_+)((r-r_-))}{(r^2-r_0^2)}}}{(r-r_+)(r-r_-)}\\
\end{eqnarray}
Calculating residue at $r=r_+$,
\begin{eqnarray}
Im{I}=-2\pi\int_{m,Q}^{m-\omega,Q-q}\frac{(r_{+}'^2-r_{0}'^2)}{(r_{+}'-r_-')}(dm'-\frac{Q'dQ'e^{-\varphi_0}}{r_+'+r_0'})
\end{eqnarray}
where $r_i=m+\sqrt{m^2+r_0^2-(Q_e^2+Q_m^2)e^{-\varphi_0}}$ and $r_f=(m-\omega)+\sqrt{(m-\omega)^2+r_0^2-(Q-q)^2e^{-\varphi_0}}$ are 
the locations of horizons before and after charged particle emission across the black hole event horizon.
Using the expression for Hawking temperature the first law of black hole thermodynamics can be expressed as\\,

\begin{eqnarray}
\frac{(dm'-\frac{Q'dQ'e^{-\varphi_0}}{r_+'+r_0'})}{\frac{(r_+'-r_-')}{4\pi(r_+'^2-r_0'^2)}}         
=dS'
\end{eqnarray}
Equation (60) therefore can be written as
\begin{eqnarray}
Im{I}=\int\frac{-1}{2}dS'=-\frac{1}{2}\Delta S
\end{eqnarray}
In this case also using the area-entropy law we find,  
$S=\pi(r_+^2-r_0^2)$,$r_0=\frac{Q_e^2e^{-\varphi_0}}{2m}$,
the change in entropy in dragged frame therefore can be expressed as 
\begin{eqnarray}
 dS'=8\pi m'[dm'-\frac{Q_e'dQ_e'e^{-\varphi_0}}{r_+'}]
\end{eqnarray}
The tunneling rate therefore turns out to be,
\begin{eqnarray}
\Gamma\approx e^{\Delta S_{bh}}
\end{eqnarray}
Thus the massive charge particle tunneling probability across dilaton-axion black hole horizon is consistent
with the theory as developed by Parikh and Wilczek. \\
As before for massless, uncharged particle the expression of $\dot{r}$ can be determined as
\begin{eqnarray}
\dot{r}=1-\sqrt{1-\frac{(r-r_+)(r-r_-)}{(r^2-r_0^2)}}
\end{eqnarray}
Substituting equation (65) in equation (41) and using first law of thermodynamics :
\begin{eqnarray}
\frac{dm'}{\frac{(r_+'-r_-')}{4\pi(r_+'^2-r_0'^2)}}         
=dS'
\end{eqnarray}
the tunneling rate turns out to be,
\begin{eqnarray}
\Gamma\approx e^{\Delta S_{bh}}
\end{eqnarray}
Once again we calculate the correction to the coefficient of the area-entropy law as follows.
The arbitrary constant $S_0$ can be found as,
\begin{eqnarray}
S_0 = S_{BH}-\pi(r_{+ in}^2-r_{0 in}^2)
\end{eqnarray}
The expression of $S_{bh}(\omega,m)$ in this case can be written as,
\begin{eqnarray}
S_{bh}(\omega,m)=S_{BH}+[\pi(r_{+ out}^2-r_{0 out}^2)-\pi(r_{+ in}^2-r_{0 in}^2)]
\end{eqnarray}
Expanding $r_{out}$ and $r_{in}$  (as in case I), we obtain,
\begin{eqnarray}
r_{+ out}=r_{+ in}(1-\epsilon\omega)
\end{eqnarray}
where $\epsilon=\frac{4m}{(4m^2-Q_e^2e^{-\varphi_0})}$\\
Now substituting $r_{out}$ and keeping terms upto first order in $\omega$ we get,
\begin{eqnarray}
S_{bh}(\omega,m)=S_{BH}(1-2\epsilon\omega)
\end{eqnarray}
Thus the modified entropy takes the form,
\begin{eqnarray}
S_{bh}(\omega,m)= \alpha_2 S_{BH}=\alpha_2\frac{A_{{BH}}}{4}
\end{eqnarray}
where $\alpha_2=(1-2\epsilon\omega)<1$\\
Therefore just as in the previous case, in this case also
\begin{eqnarray}
S_{bh}(\omega,m)\leq S_{BH}
\end{eqnarray}
Once again the entropy of this kind of black holes turns out to be less than $ S_{BH} =\frac{A_{BH}}{4}$.\\
Finally we explore the nature of the  emission spectrum of charged particle tunneling across the black hole horizons.
For Case I and Case II, expanding only up to the leading order 
the expressions for the change of entropy from equation (51) and (63) are respectively obtained as,
\begin{eqnarray}
\Delta S= -\beta[\omega-\frac{Q e^{-a\varphi_0}}{2m_0}q)]
\end{eqnarray}
and
\begin{eqnarray}
\Delta S=-\beta[(\omega-\frac{Q_e e^{-\varphi_0}}{2m}q)]
\end{eqnarray}
where $dm=-\omega$ and $dQ=-q$.The inverse temperature of these black holes 
respectively are given by $4\pi r_+(1-\frac{2r_0}{r_+})^{\frac{(1-a^2)}{(1+a^2)}}$ and $8\pi m$. 
Considering only the leading order terms in $\omega$ and $q$, the emission spectrum characteristics closely resembles to purely
thermal spectrum and can be expressed for the two cases as,\\
\begin{eqnarray}
exp[-4\pi r_+(1-\frac{2r_0}{r_+})^{\frac{(1-a^2)}{(1+a^2)}}(\omega-\frac{Q e^{-a\varphi_0}}{2m_0}q)]
\end{eqnarray}
and
\begin{eqnarray}
exp[-8\pi m(\omega-\frac{Q_e e^{-\varphi_0}}{2m}q)]
\end{eqnarray}
These can be written as $exp[-\beta(\omega-\omega_0)]$ 
where $\omega_0=\frac{Q e^{-a\varphi_0}}{2m_0}q$ for Case I and  $\omega_0=\frac{Q_e e^{-\varphi_0}}{2m}q$ for Case II.
This manifestly thermal character in the leading order of the emission spectrum is very similar to that obtained in \cite{cheng}.
However if one considers higher order, the nature of emission spectrum deviates from purely thermal spectrum. \\

{{\Large {\bf Discussion:}}\\

In the present work the expressions for tunneling 
rates of massive as well as massless uncharged and charged particles across the 
spherically symmetric Einstein-Maxwell-Dilaton-Axion black hole horizon following Parikh-Wilczek formalism have been obtained.
Earlier a similar analysis has been done in the context of a general spherically symmetric metric including the  Schwarzschild 
black hole in a noncommutative space \cite{ra}. Here we have considered spherically symmetric black hole
solutions with arbitrary dilaton-electromagnetic  and axion-electromagnetic  couplings which are specially relevant 
in string inspired models. 
We have considered both massless uncharged and and massive charged particle tunneling across  dilaton-axion coupled blackholes
having arbitrary coupling parameters for dilaton and axion with Maxwell field. The behaviour of 
massless particle is described by null geodesic as they tunnel across the horizon whereas massive particle's trajectory
does not follow null geodesic. It is well-known that the group velocity and phase velocity of massive particle can be related as,
$v_p=\frac{1}{2}v_g$. In this paper we have found out the expressions of $\dot{r}$ for both massive and massless particles
for different black holes  depending on various relations among coupling parameters a and b.
The expressions of emission rates are consistent with unitarity and provide the same results for 
both massless and massive cases. The results thus predict that 
emission rates are same irrespective of the nature of the emitted particles. In this context it can be mentioned
that since tunneling formalism depends on black hole thermodynamics where thermodynamical laws are independent of type of
the tunneling 
particles, it is expected that the emission rates are independent of the characteristics of the 
emitted particle \cite{hawk1,zh1,zh2,hu2,ya1}. Our result is a vinidication of this expectation for a much general class of 
dilaton-axion coupled spherically symmetric black holes. 

For charged particle tunneling 
we have considered two distinct situations where in case - I, the dilaton and axion coupling parameters $a$ and $b$ are comparable 
and in case - II, the axion coupling parameter $b$ is much smaller than the parameter $a$. We take into account the particle's self gravity
and also assume that the background spacetime is dynamic where energy conservation as well as electric charge conservation hold for
the whole system. Using Ist law of black hole thermodynamics we have shown that 
tunneling rate in each case is proportional to a phase space factor that depends on 
initial and final entropy difference of the black hole. Our findings are in complete
agreement with those obtained in the context of different classes of black holes and is a further validation of 
Parikh-Wilczek tunneling formalism to study Hawking radiation for a very wide class of black holes.
This work thus reveals that while the expressions for entropy, temperature etc. for the rotating Einstein-Maxwell-Dilaton-Axion black holes
considered in \cite{chen} are quite distinct from that obtained for spherically symmetric  Einstein-Maxwell-Dilaton-Axion black holes
with generalized couplings \cite{sur}, the relation between the tunneling rate and entropy change as 
obtained by Parikh and Wilczek is a very generic feature for different black hole solutions.\\           
We have also found the correction to the proportionality constant which relates the area and entropy for different cases.
It turns out that in every case its value is less than $1/4$ as proposed in Bekenstein-Hawking area-entropy law.
The modification arises from the scalar couplings in this modified class of spherically symmetric black hole solutions.   
In the leading order approximation of the entropy change the emission
spectrum exhibits thermal character. However
if one considers the entire expression without resorting to any approximation, the character of the emission is not 
thermal. This is due to the fact that the formalism respects the conservation laws and is consistent with an underlying
unitary theory with no loss of information.
Though a complete analysis of Hawking radiation through black hole needs 
quantum gravity scenarios\cite{noz1,bon,po,vor}, here 
we have used semi-classical analysis of 
Parikh and Wilczek to compute the particle's tunneling 
rate across a very general class of spherically symmetric dilaton-axion black holes with different vaues of the coupling parameters.

\end{document}